\documentclass[fleqn,10pt]{wlscirep}

\usepackage{cite}
\usepackage{graphicx}
\usepackage{amsmath}
\usepackage{amsfonts}
\usepackage{algorithmic}
\usepackage{array}
\usepackage{multirow}
\usepackage{hyperref}
\usepackage{url}
\usepackage{color}
\usepackage[symbol]{footmisc}

\newcommand{\CMPT}{\textsc{Cmpt}}
\newcommand{\CMDA}{\textsc{Cmda}}

\title{A Test for Shared Patterns \\ in Cross-modal Brain Activation Analysis}

\author[1,2,*, \dag, \ddag]{Elena Kalinina}
\author[3, \dag, **]{Fabian Pedregosa}
\author[2]{Vittorio Iacovella}
\author[1,2]{Emanuele Olivetti}
\author[1,2]{Paolo Avesani}
\affil[1]{NeuroInformatics Laboratory (NILab), Bruno Kessler Foundation, Trento, Italy}
\affil[2]{Centro Interdipartimentale Mente e Cervello (CIMeC), University of Trento, Italy}
\affil[3]{Department of Electrical Engineering, UC Berkeley, California}
\affil[*]{e\_kalinina@outlook.it}
\affil[ \dag ]{These authors contributed equally to the paper.}
\affil[ \ddag ]{Work done while PhD student at NeuroInformatics Laboratory (NILab) and Centro Interdipartimentale Mente e Cervello (CIMeC).}
\affil[ ** ]{Work done while post-doctoral researcher at the Department of Electrical Engineering, UC Berkeley.}

\begin{abstract}
 Determining the extent to which different cognitive modalities (understood here as the set of cognitive processes underlying the elaboration of a stimulus by the brain) rely on overlapping neural representations is a fundamental issue in cognitive neuroscience. In the last decade, the identification of shared activity patterns has been mostly framed as a supervised learning problem. For instance, a classifier is trained to discriminate categories  (e.g. faces vs. houses) in modality I (e.g. perception) and tested on  the same categories in modality II (e.g. imagery). This type of analysis is often referred to as cross-modal decoding.  In this paper we take a different approach and instead formulate the problem of assessing shared patterns across modalities within the framework of statistical hypothesis testing. We propose both an appropriate test statistic and a scheme based on permutation testing to compute the significance of this test while making only minimal distributional assumption. We denote this test cross-modal permutation test (\CMPT). We also provide empirical evidence on synthetic datasets that our approach has greater statistical power than the cross-modal decoding method while maintaining low Type I errors (rejecting a true null hypothesis). We compare both approaches on an fMRI dataset with three different cognitive modalities (perception, imagery, visual search). Finally, we show how \CMPT\ can be combined with Searchlight analysis to explore spatial distribution of shared activity patterns.
\end{abstract}

\begin{document}

\flushbottom
\maketitle
%
%
\thispagestyle{empty}

\section*{Introduction}%
\label{sec:introduction}

Functional MRI recordings enable the investigation of activation
patterns that characterize the working brain. The main goal is to
detect whether the neural pattern of a region of interest
correlates with a cognitive task, like, for example, object category identification.
%
Such investigations are usually focused on a specific cognitive modality, i.e.: \emph{visual perception}, \emph{real auditory}, \emph{visual imagery}, \emph{auditory imagery}, etc. A qualitative discrimination task can be designed to extract relevant information from patterns activated by two (or more) stimuli categories (like \emph{body} and \emph{car}) in one of these cognitive modalities.

Identification of activation patterns that are shared across modalities has
been the subject of numerous neurocognitive studies, with such modalities as mental
calculations~\cite{knops2009recruitment}, sensory/motor
stimulation~\cite{etzel2008testing} or words and
picture-viewing~\cite{shinkareva2011commonality}, to name a few. The most common approach here is to cast the problem of identifying common activation patterns as a supervised learning, or brain decoding, problem~\cite{haynes2006decoding, 10.3389/fnhum.2015.00151}.
Nastase and colleagues\cite{nastase2016} point out that successful classification in this setting allows to conclude that neural patterns elicited by relevant cognitive factors in one modality generalize accross to the patterns in the other modality.

In other words, a low misclassification error on a modality which is different from the one used to train the classifier provides empirical evidence that a given region of interest is involved in the cognitive
task encoded in both modalities. In the literature, such approach is referred to as cross-modal decoding analysis (\CMDA). Statistical significance of its result can be assessed using a $t$-test on the
accuracy obtained by the classifier~\cite{majerus2016crossmodal,knops2009recruitment}, or by a
permutation test based on computing the null distribution~\cite{etzel2008testing,vetter2014decoding,kaiser2016shape} of such statistic.
However, \CMDA\ suffers from a number of practical issues. As we are going to discuss in section Methods, the accuracy of a decoding model is often low in a cross-modal setting, which most probably implies an exaggerated amount of Type II errors (failure to reject the null hypothesis).


Neuroscientific investigations into the activity patterns in the fMRI data studies can be formulated as ``confirmatory'' or  ``exploratory'' analysis. Confirmatory analysis is centered on a pre-established region of interest (ROI). The \CMDA\ method presented above is often used for the confirmatory approach, when it is run on the data coming from a predefined ROI (or a set of predefined ROIs).  Exploratory analysis aims at localization of areas containing information about the presented stimuli. Here, \CMDA\ is employed in conjunction with Searchlight technique to explore the spatial structure of cross-modal activations\cite{nastase2016}. The outcome of the Searchlight procedure are maps, where each voxel is assigned some quantitative measure of information that the voxel contains about the stimulus.  The procedure of obtaining the maps consists in applying decoding sphere by sphere on time series extracted from the sphere voxels, and most commonly used information measure to produce Searchlight maps is classification accuracy. Maps are first calculated individually for each subject and then pooled together to run group analysis, where the significance of the obtained values is typically established  running $t$-tests voxelwise with respect to chance level.
Classification accuracy and $t$-tests have been subject to numerous criticisms  with regard to their role in Searchlight analysis~\cite{Etzel2013Searchlight}. Besides, use of Searchlight for cross-modal analysis faces interpretation challenges introduced by the asymmetries both in accuracies and $p$-values when training and testing on data coming from two different cognitive modalities\cite{nastase2016}.

In this work we develop a permutation test for the investigation of shared patterns across modalities that we denote
\emph{cross-modal permutation test} (\CMPT).
This test builds on a long tradition of randomization inference in the statistics literature, which can be traced back to the first half of the 20th century\cite{fisher1935design,pitman1937significance,lehmann1949theory}. Permutation tests have recently seen renewed interest in neuroimaging\cite{nichols2002nonparametric,
eklund2016cluster,Woolrich2016,WINKLER2016502} thanks to their minimal distributional assumptions and the availability of cheap computational resources.
We provide empirical evidence on synthetic datasets that this method reduces Type II errors (failure to reject a false null hypothesis) while maintaining Type I errors comparable (incorrect rejection of the null hypothesis) with respect to \CMDA. Our results highlight particular advantages of \CMPT\  in the small sample/high dimensional regime, a setting of practical importance in neuroimaging studies.
Next, we compare \CMPT\ and \CMDA\  on an fMRI study of three cognitive modalities: visual attention, imagery and perception. We conduct confirmatory analysis comparing the performance of \CMPT\ and \CMDA\  when  identifying the presence of shared patterns within a functionally defined Region of Interest (ROI). Finally, we present the results of an exploratory analysis with Searchlight making use of the proposed \CMPT\ test for the information based mapping to explore the presence of common patterns between different modalities at the whole brain level. The use of \CMPT\ allows to overcome major methodological drawbacks that had been pointed out for Searchlight in the literature\cite{Etzel2013Searchlight}.
\section*{Methods}%
\label{sec:methods}

\subsection*{Cross-modal permutation test (\CMPT)}%
\label{subsec:test}

In this section we describe a statistical test for cross-modal activation pattern analysis that we denote \CMPT. We formulate the problem of assessing cross-modal activation as a hypothesis testing problem and propose an inference procedure for this test based on a permutation schema.

\paragraph{Setting.} We assume that the experimental task consists of two modalities (e.g., {auditory} and {visual perception}) and each image in the dataset containing an activation pattern has an associated condition $A$ or $B$ (e.g., two stimuli categories like {human body} and {car}). In total, we observe $n$ activation patterns for each modality, corresponding to the number of conditions in the experiment, where each activation image is a mean image (averaged by the number of trials in the experiment) representative of one condition, and the goal is to decide whether there is a common condition effect across the different modalities.
Let us formalize this in the language of statistical hypothesis testing.
Consider the set of pairs $Z(X, Y) = \{(X_1, Y_1), \ldots, (X_n, Y_n)\}$ sampled iid from some unknown probability distribution $P$, where $X = \{X_1, \ldots, X_n\}$ (resp. $Y = \{Y_1, \ldots, Y_n\}$) are the activation patterns corresponding to the first (resp. second) modality, and where the experimental paradigm is designed such that the $X_i$ and $Y_i$ are associated with the same condition (but different modality). Since the image pairs belong to the same condition, as long as there is a condition effect shared across modality, the sequences $X$ and $Y$ cannot be independent. We hence formulate the null hypothesis (which we want to reject) that both sequences are independent and so their joint probability distribution $P$ factorizes over their marginal:
\begin{equation}
  H_0: P = P_X \times P_Y~,\quad \text{ with $P_X$, $P_Y$ the marginal distribution of $P$.}
\end{equation}

\paragraph{Test statistic.}
Given the set of image pairs $X = \{X_1, \ldots, X_n\}$ and $Y = \{Y_1, \ldots, Y_n\}$ described in the previous paragraph, let $\mathcal{A}$ (resp. $\mathcal{B}$) be the set of indices for the $A$ (resp. $B$) category. We define $X_\mathcal{A} = \frac{1}{|\mathcal{A}|}\sum_{a \in \mathcal{A}} X_a$ (resp. $X_\mathcal{B}  = \frac{1}{|\mathcal{B}|}\sum_{b \in \mathcal{B}} X_b$) as the average of activation patterns in $X$ with index in $\mathcal{A}$ (resp. $\mathcal{B}$).
$Y_\mathcal{A}$ (resp. ${Y}_\mathcal{B}$) are defined in similar way as the average of activation patterns in $Y$ with index in $\mathcal{A}$ (resp. $\mathcal{B}$).
Note that the index set is computed from images of $X$ (and not $Y$) on both cases. This asymmetry will be useful when designing the permutation scheme.
Consider also that we have access to a similarity measure between images that we denote by $\rho$. For simplicity we will initially suppose that this measure is the Pearson correlation coefficient, although we will see later that this can be generalized to any similarity measure between images.

We now have all necessary ingredients to present the test statistic that we propose to distinguish the null hypothesis from the alternative. This test statistic has values in $[-1, 1]$ and is defined as
\begin{equation}\label{eq:teststatistic}
  T(X, Y) = \frac{1}{4}\Big(\underbrace{\rho({X}_\mathcal{A}, {Y}_\mathcal{A}) + \rho({X}_\mathcal{B}, {Y}_\mathcal{B})}_{\text{within-condition similarity}} - \underbrace{(\rho({X}_\mathcal{A}, {Y}_\mathcal{B}) + \rho({X}_\mathcal{B}, {Y}_\mathcal{A}))}_{\text{between-condition similarity}}\Big)\quad.
\end{equation}
At first, its form might seem strange. Let us give two intuitions on the form of this test statistic:
\begin{enumerate}
\item \emph{As a difference of similarities}. The test statistic can be split as a difference of two terms. The first term is the sum of similarities for images from the same condition (and different modalities), while the second term is a sum of similarities for images of different conditions (and different modalities). Hence, large values of the test statistic are achieved whenever the within-condition similarity is larger than the between-condition similarity, bringing evidence for the existence of a condition-specific activation across modalities.

\item \emph{As a singularity test}. If we compute all pairwise similarities between the images ${X}_\mathcal{A}$, ${X}_\mathcal{B}$, ${Y}_\mathcal{A}$ and ${Y}_\mathcal{B}$, we obtain 4 scalars that can be arranged in a $2$-by-$2$ matrix as follows:
\begin{equation}
\begin{bmatrix}
\rho({X}_\mathcal{A}, {Y}_\mathcal{A}) & \rho({X}_\mathcal{A}, {Y}_\mathcal{B})\\
\rho({X}_\mathcal{B}, {Y}_\mathcal{A}) & \rho({X}_\mathcal{B}, {Y}_\mathcal{B})
\end{bmatrix} .
\end{equation}
Under the null hypothesis, the samples $X$ and $Y$ are independent and so ${Y}_\mathcal{A} \approx {Y}_\mathcal{B}$ (recall that the indexing was derived from images in $X$). Whenever ${Y}_\mathcal{A} = {Y}_\mathcal{B}$ the matrix above becomes colinear. A standard way to test for colinearity is through its determinant. Computing the determinant of the above equation we obtain our test statistic (modulo the normalizing factor $\frac{1}{4}$).
\end{enumerate}

%
%

\paragraph{Statistical inference.} We will estimate the distribution of this test statistic under the null hypothesis from the sample by repeatedly computing the test statistic over a permuted version of the initial sample, a technique often known as \emph{permutation or randomization test}.
For this to be valid, it is necessary to identify the quantities that we wish to permute and verify that under the null hypothesis, all permutations yield the same sample distribution\cite{lehmann2006testing}.

Consider the sequence $X_\pi$ which results from a random reordering of the activation  images in $X$ and the
sequence of pairs $Z(X_\pi, Y)$.
Under the null hypothesis, since the probability distribution factorizes over its marginal, the permuted sequence is distributed as $P_X'\times P_Y$, where $P_X'$ is the distribution of $X_\pi$. Now, by the iid assumption made previously (which is commonplace in the context of permutation testing), this distribution is invariant to permutations, and so $P'_X = P_X$ and the condition is verified.
After computing the permuted test statistic for a large number of random permutations (typically around 10000),
the significance of this test, i.e., the probability of observing a test statistic equal or as large as the one obtained, can be computed as
\begin{equation}
  p = \frac{\text{number of times}\,\{ T(X_\pi, Y) \geq T(X, Y)\}}{ \text{number of permutations}} \quad.
\end{equation}

\paragraph{Extensions.} This test extends naturally to the setting of group analysis.
In this case, the test statistic \eqref{eq:teststatistic} can be taken as the sum over all subjects of the subject-specific test statistic.
Ideally, the same permutation should be used across subjects to obtain each value of the permuted test statistic~\cite{etzel2015mvp}. It is theoretically possible to perform a two-tailed test using this test statistic. A large negative value of the test statistic would also bring evidence to reject the null hypothesis of independence. However, since the neuroscientific interpretation of such negative values is not useful for our practical purpose, we will only use the one tailed test in this paper.

For simplicity, we have considered $\rho$ the Pearson correlation coefficient as similarity measure, but the method remains valid using any other similarity measures. The Pearson correlation, being a measure of the linear correlation, works best when the effect is (close to) linear, but other more complex similarities can be used such as
 a (negative) Malahanobis\cite{Walther2016Reliability} or Wasserstein\cite{gramfort-etal:15} distance.

\paragraph{Relationship with cross-modal decoding analysis (CMDA).} \CMDA\ can be regarded within the same hypothesis testing framework outlined before, but with a different test statistic. In \CMDA, the test statistic is the accuracy of a classifier on images from one modality when it was trained on images from the other modality.

Since both \CMDA\ and \CMPT\ follow the same permutation test approach to computing significance, both rely implicitly on a label exchangeability assumption behind the data-generating process. As we have seen in the previous subsection, a sufficient condition for this is to assume that the data we observe is sampled iid. Note that this iid assumption is on the pairs from different modalities $(X_i, Y_i)$ and also on the experimental paradigm but not on the decoding train/test split, which divides the data by modality and is obviously not iid. This is a much weaker assumption than the distributional assumptions made by traditional parametric methods, it is important to keep in mind that  permutation tests are \emph{not} fully assumption-free methods and at the bare minimum require exchangeability of the observations.

A practical difference between both approaches is that the cross-modal permutation test is symmetric with respect to modalities while brain decoding is not. That is, \CMPT\ would yield the same $p$-value regardless of the order in which the different modalities are labeled. This is not true for \CMDA, where two possible tests can be performed (train on $A$ and test on $B$ or train on $B$ and test on $A$), and both can (and typically do) yield different $p$-values.

\section*{Datasets}%
\label{sec:materials}

\subsection*{Synthetic datasets}
\label{subsec:synthetic}

We construct a synthetic dataset according to a model in which the signal is a superposition of a modality-specific effect ($M_X, X_Y$), a condition-specific effect ($C_i$) and a Gaussian noise ($\varepsilon_i$):
$$
X_i = \alpha \cdot C_i + \beta \cdot M_X + \varepsilon_i , Y_i = \alpha \cdot C_i + \beta \cdot M_Y + \varepsilon_i
$$
where $\alpha, \beta$ are scalars that regulate the amount of modality-specific and condition-specific signal in the image, respectively. We then generated a total of 20 different images according to this model, considering two different modality-specific signals and two different condition-specific signals, all of them randomly generated from a Gaussian distribution. We generate 3 versions of this dataset, one with 10 voxels, one with 100 and another one with 1000 voxels.
\subsection*{fMRI dataset}
\label{subsec:fmri}

We performed empirical analysis of the data coming from a neurocognitive study of visual attention. fMRI data were collected to investigate object categorization during preparatory activity in a visual search experiment, designed in a similar manner to the one illustrated in~\cite{peelen2009neural}.


\paragraph{Participants.}24 participants (8 male, mean age 27.1, st.dev. 4.3 years) were recruited and accessed the research facility. All participantssubjects, before starting the experiment, signed a form confirming their informed consent to participate in the experimental study. After the experiment, they received monetary compensation. Each participant was instructed in advance about undergoing 2 experimental sessions (S1, S2)  on two different days. The data on all three modalities in question (perception, imagery and visual search) were acquired in the same session, S1. Out of all 24, 22 completed a significant part (6/22) or the whole (16/22) of S1. During both S1 and S2, participants were also given other tasks, which we do not report here. Out of 16 participants that underwent the whole S1 only nine participants completed 4 runs of both perception/imagery task and visual search task (8 functional runs in total). Other participants failed to reach 4 runs at least in one of the task types. So, for the analysis we are using the data of the 9 participants that have the total of 8 functional runs each. The tasks are explained in the next section. 

\begin{figure}
    \centering
    \includegraphics[width=1.0 \linewidth]{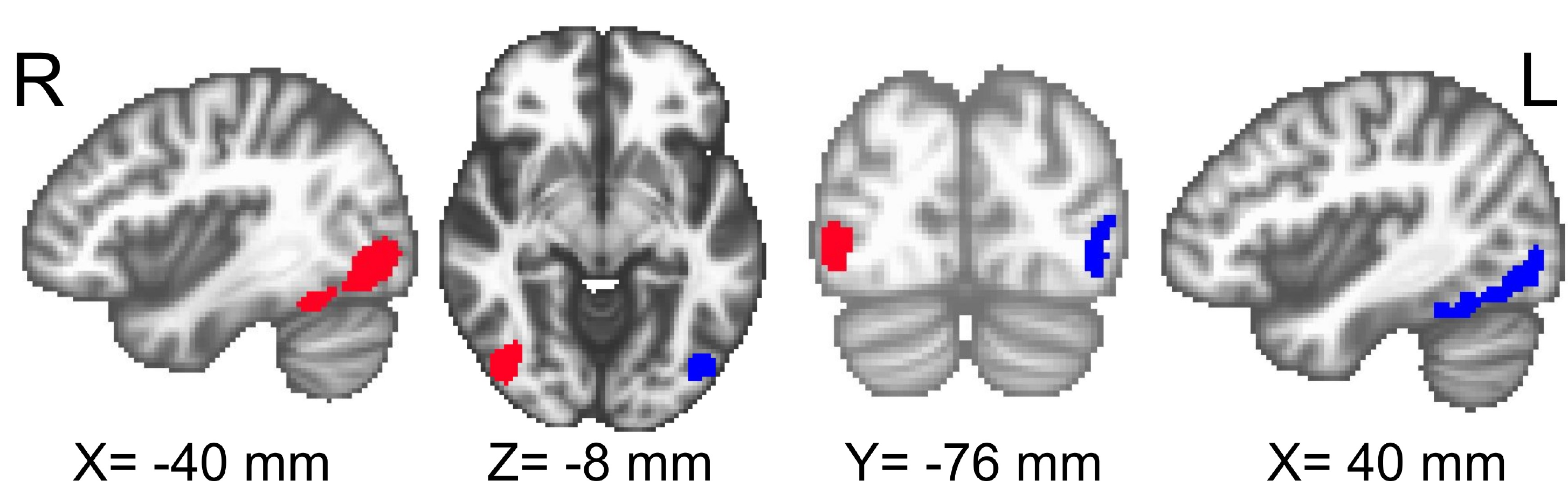}
 \caption{}
    \label{fig:roi}
\end{figure}

\paragraph{Stimuli.}Two distinctive stimulus categories were presented to the participants throughout the tasks: people (whole body image)  and cars. Participants were instructed to deal with these categories in three different ways: in \textit{perception} modality, they had to attend to 8 presentations of 16 seconds long blocks of different instances of the same category (people, here depicted by whole body figures with no face, or cars), interspersed with 16 seconds long fixation periods. Participants were equipped with a two-button box. They were requested to perform a one-back task - i.e., to  press a specific button whenever they detected the same image repeated twice in a row. In \textit{imagery} modality participants were instructed to close their eyes and mentally visualize instances of the category, indicated by the letter cue shown at the beginning of the trial. They had to press the response button whenever they achieved a mental image that was sufficiently detailed, and then they had to switch to mental visualization of another instance of the same category. At the end of the 16 seconds block, an auditory cue told the participants to open their eyes and go on with the experiment. Perception and Imagery blocks were randomly presented within the same functional run. In \textit{visual search} modality, participants were briefly (450 ms) shown images representing natural scenes (e.g.: crowded places, urban landscapes, etc ...). They were instructed through a visual cue (letter) to look for instances of one of the two categories within the scene. After scene presentation, they had 1.6 s to attend to the presentation of a mask and to give a positive or negative response by pressing a button. Visual search preparatory periods occurring between the presentation of the cue and the presentation of the scene had different lengths: 2, 4, 6, 8 or 10 seconds. Participants did not know in advance neither the lengths of preparatory period, nor their order of presentation throughout the task, which was random. Participants also had to perform a block-designed task, where we alternated presentation of images of intact and scrambled everyday objects, in order to functionally define an object selective region of interest (ROI) localized in the temporal - occipital cortex (Fig. ~\ref{fig:roi}). So, for perception and imagery the overall number of trials was 32 per modality (8 trials per 4 runs). Each visual search run consisted of 40 trials, where for each particular type of delay duration there were 8 trials. The total number of visual search trials is 160 (40 per 4 runs), while for each duration there are 32 trials (8 per 4 runs) in the dataset.  All experimental procedures had been approved by the Ethical Committee of the University of Trento and were carried out in accordance with applicable guidelines and regulations on safety and ethics. 

\paragraph{Data acquisition.} Images were acquired with a 4T Bruker (https://www.bruker.com/) scanner. For each participant, we started both experimental sessions by acquiring a structural scan using a 3D T1-weighted Magnetization Prepared RApid Gradient Echo (MPRAGE) sequence (TR/TE = 2700/4.18 ms, flip angle = $7^{\circ}$, voxel size = 1 mm isotropic, matrix = $256 \times 224$, $176$ sagittal slices). Perception / Imagery and Visual Search tasks were performed while acquiring, respectively, 177 and 195 functional scans with the following parameters: (TR/TE = 2000 / 33 ms, flip angle = $73^{\circ}$, voxel size = 3x3x3 mm, 1 mm slice spacing, matrix = 64x64, 34 axial slices covering the entire brain), during session 1. Same acquisition parameters were used for 165 scans acquired during Functional Localizer task in session 2.

\paragraph{Preprocessing.} For data  preprocessing  FSL tools were used along with in-house built Python code.
In all functional runs 5 initial volumes were discarded as dummy volumes. The skull was removed from both functional and structural images to extract the brain. Functional images were subsequently corrected for slice timing and motion artifacts. Transformation of the functional images to standard space was carried out in the following sequence. First structural scans were coregistered to the mean functional scan of each experimental run. Structural-in-functional-space images were then coregistered to standard (MNI) space to finally compute affine parameters.
To extract task-related effects from functional localizer data, beta maps for both localizer conditions (Intact vs. Scrambled objects) were computed with linear regression and next fed into contrast analysis (Intact vs. Scrambled). This analysis resulted in a ROI located in bilateral temporal occipital cortex. We selected one cluster including 625 voxels from each hemisphere, ending up with a bilateral ROI with 1250 voxels overall. We applied the ROI mask to functional data coming from Perception / Imagery and Visual Search tasks and we obtained matrices containing time-series of the ROI voxels. For CMPT analysis, 1250x16 matrices of  Perception / Imagery data and 1250x40 matrices of Visual Search data were considered per run.

\section*{Results}%
\label{sec:results}

For the rest of the paper we will refer to the rejection of the null hypothesis with the traditional significance of $0.05$ without explicitly mentioning this number.

\subsection*{Experiments on synthetic data}
\label{subsec:experiments_synth}
\begin{figure}
    \centering
    \includegraphics[width=1.0 \linewidth]{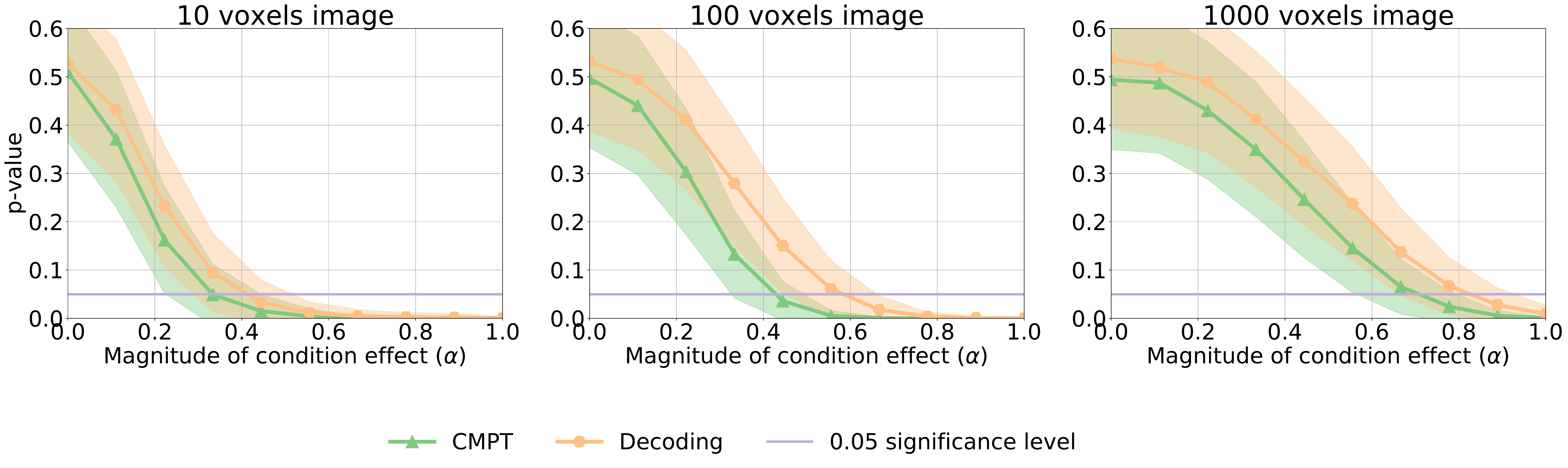}
  \caption{}
    \label{fig:simulations_1}
\end{figure}

\begin{figure}
    \centering
    \includegraphics[width=1.0 \linewidth]{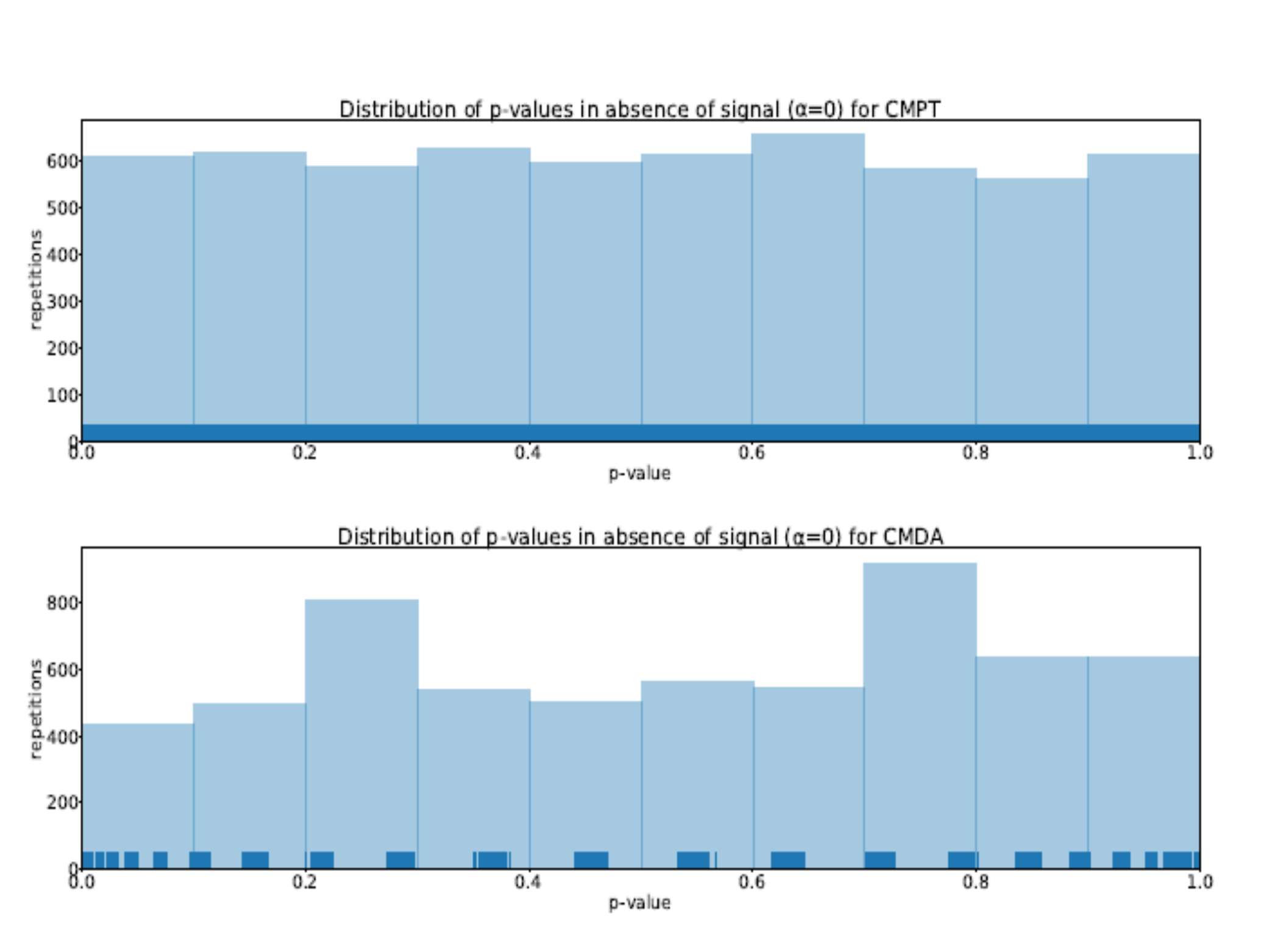}
  \caption{}
    \label{fig:simulations_2}
\end{figure}

In Figure~\ref{fig:simulations_1} we plot the resulting $p$-value after performing both \CMPT\ and \CMDA\ on the synthetic dataset described in section Methods, for varying magnitudes of the condition-specific effect ($\alpha$) and different image sizes.
In the case of decoding, this $p$-value was computed as described in \cite{ojala2010permutation}.
For $\alpha=0$, the dataset has no condition-specific signal and so the test is not expected to produce a statistically significant result. Indeed, the $p$-value of CMPT is around $0.5$. Note that because of the discreteness of the test statistic (test set accuracy), the average $p$-value need not converge towards $0.5$ as $\alpha$ goes to zero in the case for CMDA. As the magnitude of the effect ($\alpha$) increases, the method that yields a lower $p$-value has greater statistical power, because it is able to reject the null hypothesis with a greater probability.
We can see in the figure, that in general \CMDA\ $p$-values are higher, which translates into a lower probability of rejecting the null hypothesis under this approach and hence higher Type II error.

In Figure \ref{fig:simulations_2} (top row) we can see that in the absence of signal ($\alpha=0$), the distribution of $p$-values generated with \CMPT (for 6000 repetitions) is relatively flat, showing that the false positive rate (Type I error) for a significance level of $\beta$ is at the expected value of $\beta$. In the bottom row of that figure, we can see the same experiment for CMDA. In this case because of the discreteness of the test statistic, the distribution is not completely flat.

From the simulation results (Figure~\ref{fig:simulations_1}) we see that the average $p$-values yielded by \textsc{Cmpt} are always below those of \CMDA. This implies that smaller effects can be detected, and hence, that \CMPT\ has a higher sensitivity than \CMDA. Furthermore, this effect is replicated across images with different number of voxels, highlighting the benefits of \CMPT\ in the high-dimensional setting, which is of great practical importance in neuroimaging.

\subsection*{Comparison of \CMDA\ and \CMPT\ on fMRI data}
\label{subsec:experiments_fmri}
In this section we assess the agreement or disagreement in detecting shared activation patterns between \CMDA\ and \CMPT.  The similarity of activation maps for the the
discrimination of \emph{body} vs. \emph{car} categories was computed for the
following pairs of modalities: perception and imagery, imagery and visual
search. The visual search modality was investigated more in detail by first considering all durations of preparation periods put together and then
analysing separately different delays (2, 4, 6, 8 and 10s). This analysis was meant to emphasize the issue of small sample size typical for neuroscientific data. The comparison between \CMDA\ 
and \CMPT\ took into account additional elements such as the choice of ROI and
the type of encoding of activation maps.

The ROI chosen for the analysis was the Object Selective Cortex (OSC) map shown in Fig.~\ref{fig:roi}. We analysed separately the performance of methods for the left part of the ROI, right part of the ROI and the whole ROI. Encoding of the activation maps was of two types: raw BOLD and beta maps. For the raw BOLD encoding, the volumes were selected that corresponded to the peak of the hemodynamic response function (HRF, as rendered by SPM software - \url{www.fil.ion.ucl.ac.uk/spm/doc/}) convolved with the boxcar function that represented the experimental manipulation. One volume was selected per trial, and for \CMPT\ the volumes were averaged to produce a single representative volume per subject per condition per modality. For beta encoding, beta maps were calculated trialwise using linear regression; for \CMPT\ the maps were averaged over trials, too, to produce a single beta map per subject per condition per modality.

\CMDA\  was performed by training a logistic regression classifier
with $\ell_2$ regularization on the trials of one of the modalities. The
regularization parameter was selected according to a nested cross-validation
scheme (leave-one-run out). The accuracy was then estimated on a test set from another
modality. The training and test process was replicated for each
subject. Then, the $p$-value was computed for the group using the permutation scheme described in~\cite{etzel2008testing}. The resulting $p$-values are reported in
Table 1.

\CMPT\  group analysis was carried out as described in section Cross-modal permutation test (\CMPT). The similarity distance between activation
maps that we used is the Pearson correlation measure, both for raw BOLD
volume and beta maps encoding. The significance of the proposed test
statistics was computed by a permutation scheme with $10.000$
iterations to estimate the null distribution. The resulting $p$-values are
reported in Table 1.

\begin{figure}
    \centering
    \includegraphics[width=1.0 \linewidth]{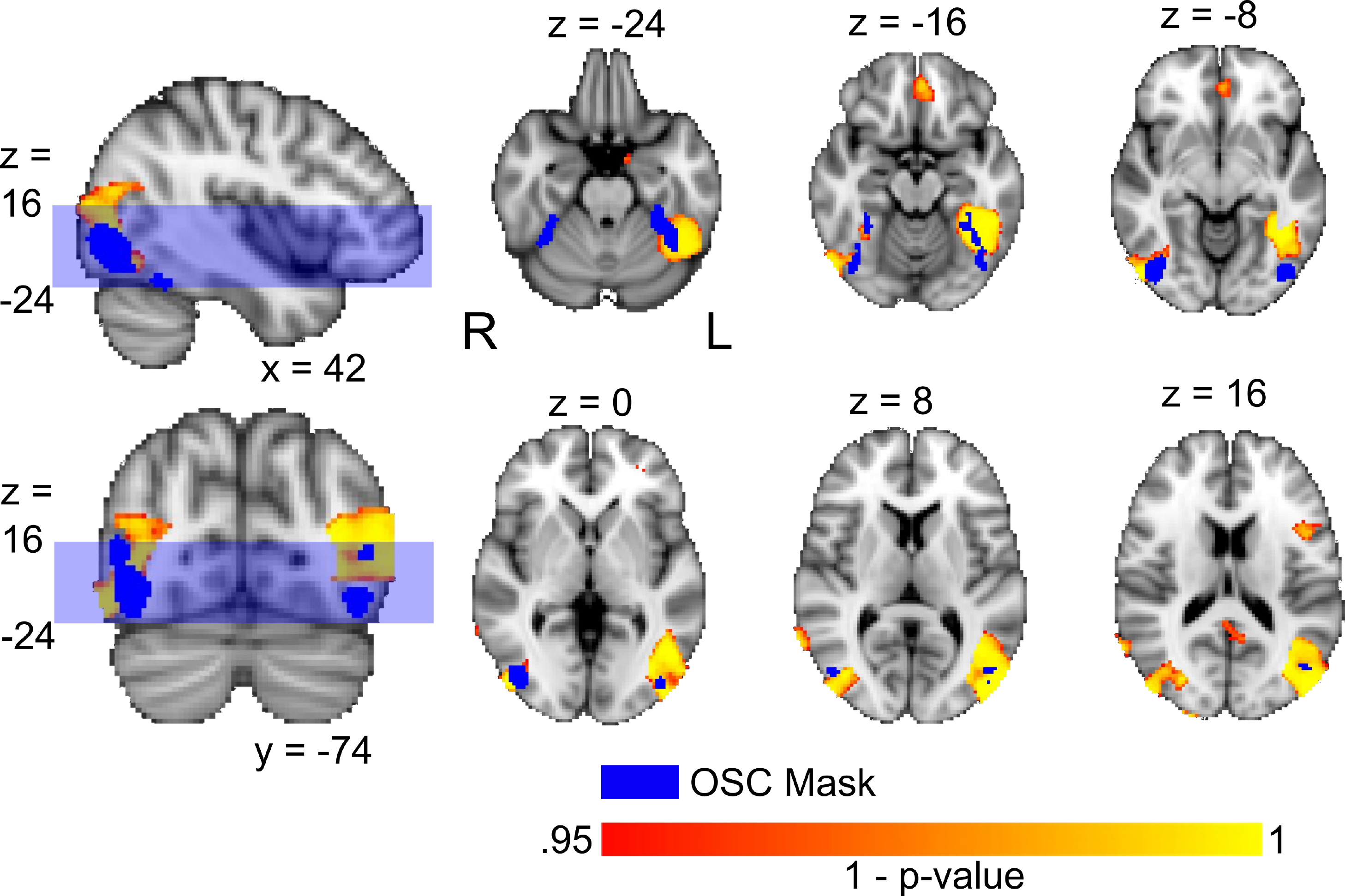}
 \caption { }
    \label{fig:mask_sl}
\end{figure}

In Table 1  we report only one result for the
comparison between \CMDA\ and \CMPT\ related to raw BOLD volume encoding: the cross-modal analysis between Perception and Imagery. In this case none of the methods detect a meaningful shared activation pattern. Beta maps encoding on the other hand seems to be a more efficient representation. Chen and colleagues \cite{Chen2006Exploring} demonstrate that using beta values is a way to get rid of intrinsic variabilities of BOLD signal throughout the brain and, specifically, within a single area. In our case, this means that beta values are more representative of the effect size than raw BOLD signal changes during task-on periods. For this reason in the presentation of results we only focus on results obtained with data encoded with beta maps.

The results in Table 1 confirm our expectations about the presence of common patterns between modalities in Object Selective Cortex. At the same time, \CMPT\ appears to have higher statistical power and sensitivity in revealing these patterns. The results reported in Table 1  illustrate two main scenarios: both \CMDA\ and \CMPT\ show significant $p$-values or only \CMPT. In light of the simulation results we may argue, that since \CMDA\ has a higher false error rate  (Figure~\ref{fig:simulations_1}), in case of such disagreements the \CMPT\ result is more reliable. This argument is further supported by the additional empirical evidence that the false positive rate or Type I error is similar for the two tests, limiting the risk of the disagreement being biased by a more optimistic rejection of the null hypothesis.

Cross-modal analysis results for perception vs. imagery with beta maps encoding are in agreement between \CMDA\ and \CMPT\ when the two hemispheres are considered individually, namely the left and right OSC respectively. When the analysis is extended to the joint ROI, the number of trials remains constant, while the number of voxels double. In this case the classifier is affected by the higher dimensionality of data, and \CMDA\ does not succeed in rejecting the null hypothesis.

The empirical results also support the claim that \CMPT\ is more robust not only in high-dimensional but also in small sample setting. Simulations show that the Type II error of \CMPT\ is below that of \CMDA\ in the small sample regime  (Fig. ~\ref{fig:simulations_1}). We may find analogous behaviour for the cross-modal analysis of visual search vs. imagery. If we consider the cumulative trials of visual search, irrespective of the delays, \CMDA\ rejects the null hypothesis. When we restrict the cross-modal analysis to single delays of preparation period for visual search, the number of trials drop from $160$ to $32$. In this case \CMDA\ fails to reject the null hypothesis while \CMPT\ does not.

\CMPT\ results are in line with the view that we should expect the presence of shared activity patterns between perceived and imagined object categories. \CMPT\ analysis also confirms that the presence of these patterns can be expected in high-level visual areas processing information about object categories. We are going to further elaborate on this point in the discussion section. On the other hand, \CMDA\ results appear to be affected by the data sample size relative to the high dimensionality of data.

\subsection*{Exploratory data analysis with \CMPT}%
\label{subsec:exploratory_CMPT}

We ran Searchlight analysis of the whole brain, using
\CMPT. First, we wanted to show if and how inserting \CMPT\ as the
elementary unit within the Searchlight framework could identify voxels that
store information related to common activation patterns for two
different cognitive modalities. Next, we intended to compare the spatial
profiles of the exploratory analysis with the
ROI individuated for the confirmatory
analysis. 

To construct group level maps, we referred to the procedure
we illustrated in section Cross-modal Permutation Test (\CMPT) at a ROI level. Here we
consider  the spheres centered on each voxel as ROIs: we first compute the
"true" statistic and then we proceeded by using permutations. We started by computing single-participants' \CMPT\ - Searchlight maps, where
each voxel was considered as the center of a sphere (r= 8), and
calculated the T-statistic. Then, we summed up single-participants'
T-values and ended up with the true group-level statistic. Next, we
created N=10000 permutations of the session labels, and we subsequently
constructed 10000 averaged beta maps for each of the conditions based
on the permuted labels - that is, two maps per subject per modality per
permutation. In this way, we made sure that the data coming from different
participants were tested against the same permutations in a uniform
way. We then applied \CMPT\ procedure on these permuted maps by first computing an
individual T-statistic and then by summing up the group values - that is,  we constructed an ad-hoc null distribution. Finally, we simply
counted how many times the ``true'' group statistic was higher than the
permuted group statistic, and we transformed the count in a fraction of
the total number of permutations, obtaining a p-value. This value was
assigned to the voxel in the center of the sphere. The procedure was
repeated for each voxel within the gray matter mask. We ended up with a \CMPT\- Searchlight map of p-values coming from a combination
of permutation-like tests.

\begin{figure}
        \centering
    \includegraphics[width=1.0 \linewidth]{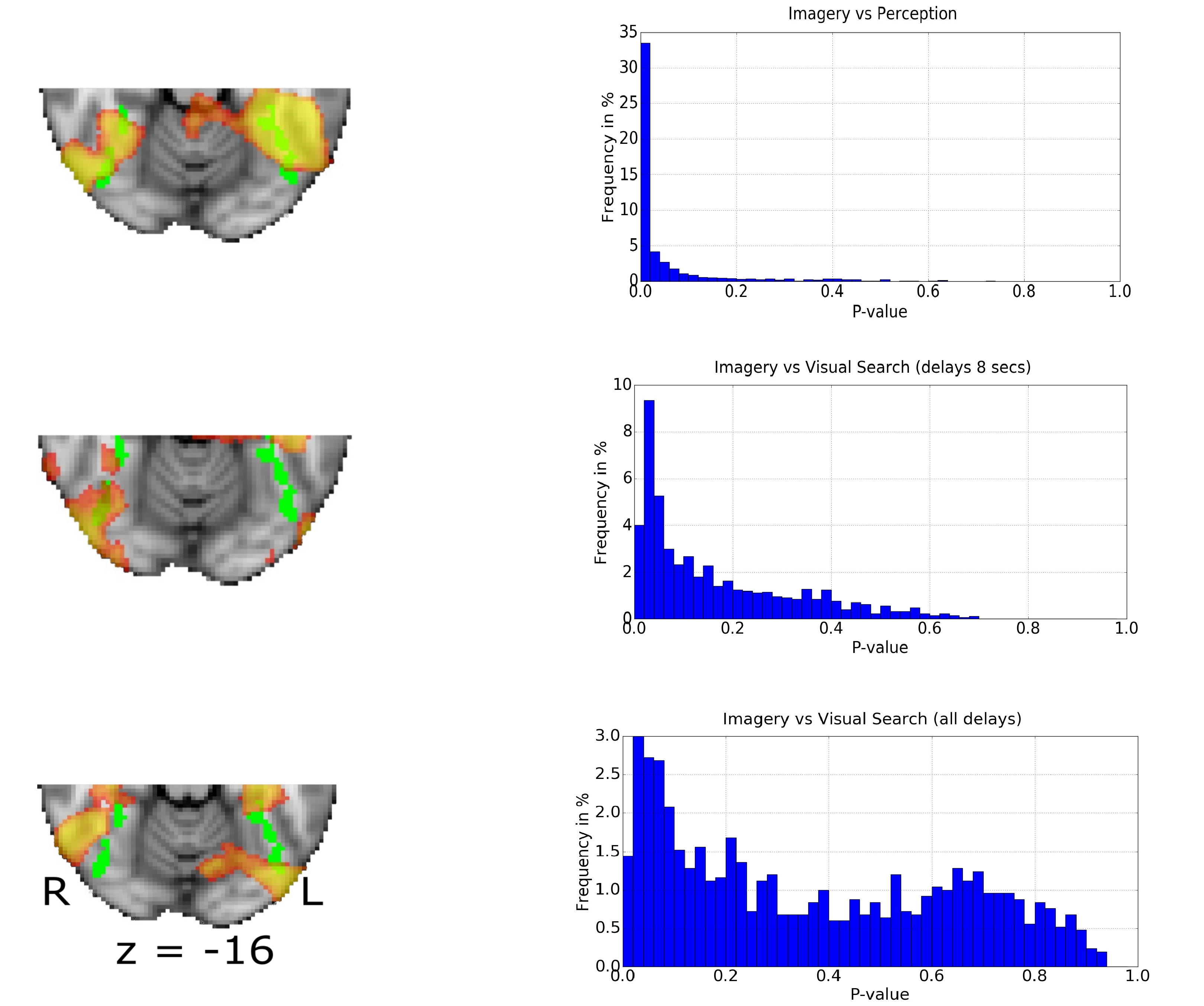}
\caption{}
\label{fig:results2}
\end{figure}

In Figures ~\ref{fig:mask_sl} and  ~\ref{fig:results2} we present results coming from the
exploratory analysis. Fig. ~\ref{fig:mask_sl} showcases the overlap between the whole OSC ROI and informative voxels identified by Searchlight in the occipital-temporal cortex for the cross-modal pair of perception vs. imagery only. In Fig. ~\ref{fig:results2}, we put together fragments of maps for the pairs of cognitive
processes where our confirmatory analysis yielded significant results,
namely perception vs. imagery, visual search (delay 8s)
vs. imagery, visual search (all delays) vs. imagery (see Table 1~\ref{tab:results1}). For merely illustrative purposes, the maps were thresholded at
the conventional significance level of 0.05 (as we are not aiming at
significant cluster identification, no correction for multiple
comparisons was carried out). In the left column of
Fig. ~\ref{fig:results2} we demonstrate the overlaps between the ROI
identified in the course of the group analysis  (Fig. ~\ref{fig:roi}) and the
portions of the map that signal the presence of information about
common patterns between two cognitive processes for a single slice (z=-16). The fact that the ROI
identified contains a high portion of informative voxels is further
illustrated by the histograms in the right column of the same
figure. These are histograms of the p-values of the voxels within the
ROI. We can see that all three histograms have a skewed shape,
signalling the presence of a rather large number of voxels with
p-values under 0.05 in the ROI.

For comparison, we also ran \CMDA\ Searchlight with the same sphere size (r=8) for the same modality pairs: perception vs. imagery, visual search (all delays) vs. imagery, visual search (delay 8) vs. imagery. The analysis was performed with Matlab 8.5.0, MathWorks, NatickMA, USA using in-house code and Libsvm library  (https://www.csie.ntu.edu.tw/~cjlin/libsvm/). Classifier used for producing the maps was an SVM classifier with a linear kernel as implemented in the Libsvm library. For each subject, two Searchlight maps were obtained for each modality pair, one where the classifier was trained on the Imagery data and tested on the other modality data, and one where the assignment of train - test data was reversed. Then, these two maps were averaged as suggested in\cite{nastase2016} yielding a single map per subject per modality pair. For the group analysis, a one-sample t-test against chance level (50 \%) was performed using SPM software. The resulting group maps were thresholded at the significance level of 0.05 and cluster size of 10 voxels. Then we compared the group maps to the OSC ROI selected for the confirmatory analysis. The results are presented in Table 2. It shows percentages of voxels within the OSC ROI that were identified by the \CMDA\ Searchlight as informative about shared patterns between two modalities. The numbers concerning the size of intersection between the Searchlight map and the ROI are given both as an absolute number of voxels within the ROI and in terms of percentages. 

The problem of asymmetry between classifier results when swapping train and test modalities in a cross-modal setting is well attested for the pair of perception vs. imagery. The accuracies obtained with the classifier trained on the imagery data have been shown to be consistently higher than after training on perception data (~\cite{Reddy2009category, cichy2012imagery}). To minimize the impact of this asymmetry in cross-modal investigations,  it was suggested to average the maps resulting from different train-test combinations for a pair of modalities\cite{nastase2016}. However, the authors of the paper showed that the divergence in accuracy numbers obtained with different train-test combinations in their data was insignificant. 

\CMDA\ Searchlight results in our dataset seem to be rather seriously affected by the issues stemming from the accuracy asymmetries. First, for the pair of perception vs. imagery, \CMDA\  Searchlight trained on imagery data  identifies a high number of voxels within the OSC mask, both in right and left OSC. If trained on perception data, the Searchlight finds a much lower number of voxels within the same area, all of them in the right OSC. In the averaged mask, the number of the voxels that survive is nearly 10 times lower than that identified by the Searchlight trained on imagery data (49 against 432). This same kind of asymmetry is even more prominent for the pair of imagery vs. visual search:  Searchlight, trained on imagery data, identifies voxels within the OSC ROI, while it does not identify any in the same area if trained with visual search data (neither all delays, nor delay 8). This result makes us pose a question about the extent to which the voxels idenitified belong actually to really shared patterns between modalities, or we should rather talk about voxels in one modality that are informative about the patterns in the other modality.  Our overall conclusion about  \CMDA\  Searcchlight is that its use might be questionable in cases when notable asymmetry is expected, as is the case with perception vs. imagery. For some modality pairs asymmetry does not seem to be a big  issue, as is the case with the data used in \cite{nastase2016}, and the use of \CMDA\ Searchlight could be more justified for these data. 

\section*{Discussion}%
\label{sec:discussion}

The patterns of brain activity that are shared between the cognitive processes of perception and imagery  have been the subject of quite numerous studies. The question investigated was if we can arrive at abstract, top-down object representations~\cite{Reddy2009category, Ishai2010imagery} containing distinguishing features~\cite{Roldan2017Object, Horikawa2017Generic} that will have common neural substrate both for viewed and imagined object categories~\cite{Roldan2017Object}. To test for the presence of shared patterns, many studies used cross-modal decoding - namely, multivariate pattern analysis with SVM classifiers~\cite{Lee2012Disentangling, Reddy2009category, cichy2012imagery}. Significant cross-modal classification accuracies were taken as the evidence in favour of the presence of shared activity patterns. In some studies, correlation-based analysis was also performed to visualize and estimate similarity between these patterns in terms of distance ~\cite{Lee2012Disentangling,Reddy2009category}. What emerged from these studies was the view that, indeed, visual imagery activates the same areas that contain information about visually perceived stimuli ~\cite{Farah1989Neural, Roldan2017Object}, and shared patterns for stimulus categories in these two processes can be established~\cite{Pearson2015Mental,Lee2012Disentangling, Reddy2009category,Stokes2009Topdown,cichy2012imagery, Anderson2015Reading}. The areas where these common representations were found include the ventral temporal pathway,  lateral occipital cortex ~\cite{Stokes2009Topdown, Lee2012Disentangling, Reddy2009category, cichy2012imagery} and extrastriate cortex~\cite{Lee2012Disentangling, Pearson2015Mental, Stokes2009Topdown, Ishai2010imagery}. The question of shared patterns in early visual areas, such as V1, remains controversial ~\cite{Lee2012Disentangling, Reddy2009category}.  Horikawa~\cite{Horikawa2017Generic} showed that it depended on the feature type: lower visual features had similar representations for perception and imagery in lower visual areas, while the same was true for higher visual features in higher visual areas. Cichy~\cite{cichy2012imagery} arrived at a similar conclusion about the subdivision of features: although they did not find significant accuracies for decoding object categories in lateral early visual cortex, they could identify shared representations of object locations in these areas.

Top-down attention patterns mediate attention biases during perception and affect behavioural performance in attention related tasks. In case of visual attention, these patterns can be revealed in visual search experiments via activity in the category-related object selective areas during preparatory delays~\cite{Stokes2009Shapespecific}. Several studies attempted at demonstrating the  high-level nature of the preparatory patterns through cross-modal analysis, mostly with visual perception as the other modality ~\cite{Stokes2009Shapespecific, peelen2009basis, peelen2009neural}. As object representations in the brain obtained during imagery tasks are thought to be closer to high-level top-down representations of objects in visual cortex ~\cite{Pearson2015Mental, Roldan2017Object}, the hypothesis naturally suggests itself that we can expect these patterns to show up also during visual search preparatory periods. We tried to shed light on this hypothesis using both \CMDA\ and \CMPT\ on visual imagery and visual search data. Besides, we ran cross-modal analysis separately for preparatory periods of varying length (between 2 and 10 seconds) to get insights into preparatory dynamics. We were expecting that only certain dealys would result significant, conforming different hypotheses about this dynamics. For instance, if only shorter delays (2-4 s) had resulted significant, that could be evidence in favour of transitory and cue-related nature of the preparatory activity in the Object Selective Cortex. If, on the other hand, we had seen significant results in the longer delays, that could reveal the fact that it takes time for the activity to build up.  First, we see that both methods confirm expectations about imagery patterns being more high level than perception. None of the methods yielded significant results in the pairs of visual search vs. perception. As for the presence of the shared patterns between visual imagery and visual search, we are faced again with limitations of \CMDA\ as a method: its results can be significant and it can reveal the presence of shared patterns between preparatory periods and imagery, but this type of analysis needs a lot of data. On the other hand, \CMPT\ can reveal shared patterns even with fewer data as is the case with 8 seconds delay. Further study is needed to uncover the temporal dynamics of the preparatory top-down patterns. We hypothesize that delays shorter that 8 s do not allow the preparatory activity to build up, while in case of 10 s the delay  it is too long, and the subject might be loosing concentration after a certain period of time.

We placed \CMPT\ side by side with other standard data analysis techniques in order to examine whether this approach could be as informative as others. We have shown that
in confirmatory, top-down contexts \CMPT\ can yield better results than \CMDA. However, it is necessary to mention one limitation of the method. One of the overarching questions in the study of visual imagery is identifying neural representations of categorical features in the form of brain activation maps~\cite{Pearson2015Mental, Roldan2017Object}.  \CMPT\ method cannot provide insights into the location of the discriminative patterns at a ROI level.  Despite being a more robust test for cross-modal analysis, \CMPT\ is not appropriate to investigate the shape of shared pattern within a given ROI. In this case, \CMPT\ doesn't support a sensitivity analysis at the voxel level needed to compute granular brain maps of activations that are common between modalities.  On the other hand, \CMDA\ (at least when linear classifiers are used) contains a vector of weights that can give some clues about the relevance of the input features. However, \CMPT\ combined with Searchlight technique can be a helpful method to locate brain regions that contain information about common patterns between modalities.

We took advantage of one strong point of the Searchlight analysis, its ``modular'' nature: Searchlight might be thought of as a generic framework of data examination that can subsume various analysis techniques as elementary units. Searchlight is widely used in neuroimaging, but it suffers from a number of issues. Conducting Searchlight analysis has several major advantages: first, it can be run on the whole brain, no prior ROI selection is required. Next, it avoids the ``curse of dimensionality'' of full brain classification, by reducing the number of features used at each point by the classifier. Finally, it has proven to be quite successful in identifying subject specific activation patterns~\cite{Etzel2013Searchlight}. The maps produced with Searchlight are of the same nature as the maps obtained with the univariate GLM approach, but they are based on a more fine-grained pattern identification from multiple voxels and better reflect the spatial properties of the BOLD signal (that is, adjacent voxels have similar activation patterns). However, major criticisms of the Searchlight approach regard the use of classification accuracy as the information measure and the t-test as the method to obtain group significances. As is pointed out in~\cite{Etzel2013Searchlight}, SVM classifiers can correctly classify even with a few number of highly informative voxels and when weakly informative voxels are numerous enough. Both of these behaviours can cause distortions in a map: in the first case, all searchlights overlapping with one of a few informative voxels will be significant. In this way, the number of informative voxels is overestimated. In the second case, the cause of distortions is ``discontinuous information detection'': groups of weakly informative voxels will be missed out if their size is below a certain threshold, but can be judged significant if you just add a single voxel. That leads to underestimation of the number of significant clusters just because the number of weakly significant voxels does not reach a certain mass.  Efficiency of using classifiers with Searchlight depends strongly on the classifier parameters and sphere size~\cite{Etzel2013Searchlight,Allefeld2014Searchlightbased}. In \cite{Allefeld2014Searchlightbased}, the point is raised against interpretability of classifier accuracy with neuroscientific data: unlike distance measures, its value depends on the properties of the dataset (amount of training data and what kind of data is used as test data) and not only on the presence of a particular effect in the data. Besides, the authors point out that capturing interactions of several factors in a factorial experimental design cannot be cast as a classification task. So, addressing these methodological issues for Searchlight can  significantly improve this  valuable tool and make its result more scientifically rigorous.

Classification accuracy is not the only way to represent  information content. In the original paper by Kriegeskorte ~\cite{Kriegeskorte2006Informationbased} the metrics used was Mahalanobis distance between the distributions corresponding to stimulus categories. In ~\cite{Allefeld2014Searchlightbased}  the authors build on the probabilistic model of the data proposing a cross-validated multivariate ANOVA (MANOVA) as the informational content measure. In~\cite{Walther2016Reliability}  three various measures - classification accuracy, Euclidean/Mahalanobis distance, and Pearson correlation distance - are compared for reliability in the context of Searchlight analysis. In this paper, it was shown that  “continuous crossvalidated distance estimators” such as Euclidean/Mahalanobis distance or Pearson correlation should be preferred for Searchlight because they are more interpretable from the neuroscientific viewpoint.

Another bunch of critical remarks concerns the use of t-tests for assessing significance at the group level. Certain properties of neuroscientific data make the use of t-tests questionable for this purpose, “particularly, the low number of observations and the non-gaussianity of the probability distribution of accuracy. As a consequence, several assumptions of the t-statistic are not met, rendering the procedure invalid from a theoretical point of view" ~\cite{Stelzer2013Statistical}. However, it is not the only option here. In~\cite{Stelzer2013Statistical} a non-parametric test for group significance and cluster inference was proposed based on permutations and bootstrapping procedure. Nastase and colleagues \cite{nastase2016} also opt for permutation tests in Searchlight context.

Methodologically, using \CMPT\ in conjunction with the Searchlight technique for cross-modal pattern analysis has several advantages over the common Searchlight procedure because it does not rely neither on classification accuracy nor on the t-tests and hence avoids the common methodological pitfalls. At the same time, we are following the suggestions in the literature that are considered more appropriate for the Searchlight.  First, the test statistic proposed in equation ~\ref{eq:teststatistic} that is used as the measure of information contained at each voxel is based on Pearson correlation and is interpretable in terms of similarity. Second, group significance is tested non-parametrically with permutation tests that do not make assumptions about the shape of the data distribution. We found that \CMPT\ integrated into Searchlight  has proven effective also to explore and, potentially, confirm what we observed using top-down, ROI-based analysis, which suggests both robustness and efficiency of the \CMPT\ Searchlight in fMRI data analysis.

However, it is important to note that confirmatory and exploratory analyses report different p-values. While it is possible to qualitatively compare the outcomes of these two analyses, plainly putting their  p-values side by side might be misleading. CMPT-ROI p-values come from an extended, functionally well-defined area including 625 or 1250 voxels. CMPT-SL analysis spans over the whole brain sphere by sphere, extracting results from spheres including about 200 voxels each. This means that p-values coming from confirmatory and exploratory analysis should not be compared on a purely quantitative level.

The question of shared patterns between various cognitive modalities is relevant not only for object categorization in visual processing. It is fundamental in the study of interactions between top-down and bottom-up processing streams in the human brain in general. Further directions of study could include using the \CMPT\ method and the \CMPT\ Searchlight technique  with a wider number of other cognitive modalities, such as auditory or linguistic~\cite{Simanova2014Modalityindependent, Simanova2016Linguistic, Borghesani2014Perceptualtoconceptual}.  Besides, we could investigate other areas that can share representations with imagery - for instance, working memory areas~\cite{Pearson2015Mental}. Finally, the \CMPT\ method could be tried with other types of neuroimaging data - as, for example, EEG motor imagery data for Brain-Computer interfaces  ~\cite{Choi2013Electroencephalography} or MEG data ~\cite{Dikker2013Predicting, Hirschfeld2011Effects}.


\section*{Acknowledgements}
The research was partially funded by the Autonomous Province of
Trento, Call ``Grandi Progetti 2012'', project ``Characterizing and
improving brain mechanisms of attention - ATTEND'' and by Centro
Internazionale per la Ricerca Matematica (CIRM). The authors  gratefully acknowledge the contribution of Marius Peelen to the disucssions of experiment design and data analysis and to Valentina Borghesani for valuable feedback on the manuscript.

\section*{Author contributions statement}

E.K., V.I. and P.A participated in designing the experiment. E.K. and V.I. implemented the design, conducted the experiment, acquired the data and preprocessed the data. All authors analysed the data. E.K., F.P. and E.O. implemented data analysis pipelines.  F.P. designed, implemented and conducted experiments on simulated data. E.K., F.P., V.I. and P.A. wrote the manuscript.

\section*{Additional information}
The authors declare no competing interests.

\section*{Data and code availability.} Processed data (ROI beta-maps, BOLD volumes and Searchlight maps) are available from a public repository on Github \url{https://github.com/elena-kalinina/Code_CMPT_paper}. The code used to generate and analyze synthetic data is available from the same repository along with data analysis code and Searchlight implementation. Sharing the whole dataset involves the consent of third-parties who participated in the experiment design. If the consensus is reached, the data could be made publicly available.



  \begin{center}
    \begin{tabular}{|c|c|c|c|c|}
      \hline
      Modalities & ROI & Encoding & CMDA & CMPT \\
      \hline
      P vs I & L+R & Bold & 0.12  & 0.75 \\
      P vs I & L+R & Beta & 0.08 & {\bf 0.001}  \\
      P vs I & L   & Beta  & {\bf 0.001} & {\bf 0.001}  \\
      P vs I & R   & Beta & {\bf 0.005} & {\bf 0.008}  \\
      V vs I (delay 2 s) & L+R & Beta & 0.272 & 0.589 \\
      V vs I (delay 4 s) & L+R & Beta & 0.387 & 0.141 \\
      V vs I (delay 6 s) & L+R & Beta & 0.165 & 0.303 \\
      V vs I (delay 8 s) & L+R & Beta & 0.659 & {\bf 0.004} \\
      V vs I (delay 10 s) & L+R & Beta & 0.249 & 0.461 \\
      V vs I (all delays) & L+R & Beta & {\bf 0.012} & {\bf 0.001} \\
      \hline
    \end{tabular}

  \end{center}

Table 1. Comparison of cross-modal decoding analysis (\CMDA) and cross-modal permutation test (\CMPT). The values refer to p-value for the discrimination task of \emph{body} vs. \emph{car} in each pair of the cognitive  modalities listed in the extreme left column, where P stands for  \emph{perception}, I for \emph{imagery} (I) and V for \emph{visual search}.
  \label{tab:results1}
\noindent \\

\begin{center}
    \begin{tabular}{|c|c|c|c|c|}
      \hline
      Modalities & ROI & Map Type & Intersection N Vox & Intersection \% \\
      \hline
      P vs I & L &  mean &  4 & 0.64 \\
      P vs I & L & single P I & 0 & 0 \\
      P vs I & L  & single I P  & 239  &  38.24 \\
      V vs I & L  & mean & 5 &  0.8 \\
      V vs I & L & single V I & 0 & 0 \\
      V vs I & L & single I V & 12 &  1.92 \\
      V vs I (delay 8) & L & mean & 0 &  0 \\
      V vs I (delay 8) & L & single I V & 29 &  4.64 \\
      V vs I (delay 8) & L & single V I & 0 &  0 \\
      P vs I & R &  mean &  45 &  7.2 \\
      P vs I & R & single P I & 13 & 2.08 \\
      P vs I & R  & single I P  & 193  &  30.88 \\
      V vs I & R  & mean & 0 &  0 \\
      V vs I & R & single V I & 0 & 0 \\
      V vs I & R & single I V & 1 &  0.16 \\
      V vs I (delay 8) & R & mean & 0 &  0 \\
      V vs I (delay 8)& R & single I V & 10 &  1.6\\
      V vs I (delay 8)& L & single V I & 0 &  0 \\
      P vs I & L+R &  mean &  49 &  3.92 \\
      P vs I & L+R & single P I & 13 & 1.04 \\
      P vs I & L+R  & single I P  & 432  &  34.56 \\
      V vs I & L+R  & mean & 5 &  0.4 \\
      V vs I & L+R & single V I & 0 & 0 \\
      V vs I & L+R & single I V & 13 &  1.04 \\
      V vs I (delay 8) & L+R & mean & 0 &  0 \\
      V vs I (delay 8)& L+R & single I V & 39 &  3.12 \\
      V vs I (delay 8)& L+R & single V I & 0 &  0 \\
   
      \hline
    \end{tabular}
\label{tab:results2}. 
  \end{center}

Table 2. Percentage of voxels within the OSC ROI identified by the \CMDA\ Searchlight as informative about shared patterns between two modalities. The cognitive modalities are listed in the extreme left column. P stands for  \emph{perception}, I for \emph{imagery} and V for \emph{visual search}. In the second column, it is shown which part of the OSC mask was used for calculations: left (L), right (R) or the whole mask (L+R). Number of voxels in L+R ROI is equal to 1250; each half of it (both L and R) has 650 voxels. The third column shows which type of map was used for calculations: an averaged map (denoted as mean), or a single training - test pair map (denoted as single). In the names of the single maps, first comes the training modality, and second the test modality. In the last two columns you can find the numbers concerning the size of intersection between the map in column 3 and the ROI in column 2. In column 4, the intersection size is given as an absolute number of voxels within the ROI. In the last, extreme right column, it is given in terms of percentages.

Figure 1. Object Selective Cortex (OSC) group map in  temporal-occipital cortex delineated based on the functional localizer (intact vs. scrambled objects). Its right (R) hemisphere part is shown in red, left (L) in blue. X, Y and Z locate the coordinates of the slices.
\noindent \\

Figure 2. Statistical power as a function of the effect size. We compare the obtained $p$-value (y) for different magnitudes of the condition-specific effect (x) and different image sizes (10, 100 and 1000 voxels). Shaded areas correspond to the standard deviation.  Across these different scenarios, \CMPT\ yields a higher significance than the decoding-based approach.
\noindent \\

Figure 3. The distribution of $p$-values generated with \CMPT\  (top) and \CMDA\  (bottom) in the absense of condition-specific effect after 6000 repetitions of the experiment. $p$-values are on the $x$ axis, while the frequencies are shown on the $y$ axis.  In case of \CMPT\  (top) the distribution converges to the uniform. The false positive rate (Type I error) for a significance level of $\beta$ is at the expected value of $\beta$. For \CMDA\  (bottom), the distribution is not fully flat due to the discreteness of the scoring rule.

Figure 4. Overlap between the whole OSC ROI and informative voxels identified by Searchlight in the occipital-temporal cortex for the cross-modal pair of perception vs. imagery. The ROI (the same as in Figure {fig:roi}) is shown in blue. P-values in the Searchlight map are colorcoded. The maps were thresholded at the conventional significance level of 0.05, and the color code shows values between 0.95 (corresponding to the p-value of 0.05) in red and 1 (corresponding to the p-value of 0.00) in yellow. The leftmost column of the figure shows sagitarial (top) and coronal (bottom) slices (at x= 42 and y = -74 correspondingly) where the rectangle shaded in blue delimits the axial slices further presented in columns 2 - 4. These are six slices of the axial plane taken at steps of 8 between coordinates z = -24 and z = 16. We can see that in all these six slices, there are overlaps between the OSC ROI and areas where Searchlight identified voxels informative about shared activity patterns between perception and imagery.
\noindent \\

Figure 5. \CMPT\ - Searchlight analysis results. The rows represent results coming from the pairs of cognitive processes where our confirmatory analysis yielded significant p-values (Table 1): perception vs. imagery (top), visual search (delay of 8 s) vs. imagery (middle), visual search (all delays) vs. imagery (bottom). In each row, the left column shows the overlap between the Object Selective Cortex ROI previously defined for confirmatory analysis (Fig. ~\ref{fig:roi}) and the Searchlight maps obtained for this pair. In the right column, there is a histogram of p-values within the OSC ROI  coming from the Searchlight map. The left column showcases axial slices of Searchlight maps for the corresponding pairs, focusing on posterior region.  Slices are taken at z= -16, maps were thresholded at the significance level of 0.05, and the color code shows values between 0.95 (corresponding to the p-value of 0.05) in red and 1 (corresponding to the p-value of 0.00) in yellow.  The OSC ROI is presented in green. Overlaps between the green area and colorcoded maps can be observed for all three pairs. The right column further illustrates the fact that the confirmatory OSC ROI identified contains a high portion of informative voxels. In each histogram, the x axis represents the p-values, while their frequencies (in \% from the overall number of voxels in the area) are ordered along the y-axis. All three histograms are skewed to the right, signalling the presence of a rather large number of voxels with p-values under 0.05 in the ROI.

\end{document}